\documentclass[11pt]{article}
\usepackage{latexsym,amssymb}
\def\1bar{1\hskip -.275cm -}
\def\2bar{2\hskip -.275cm -}
\def\3bar{3\hskip -.275cm -}

\newsavebox{\uuunit}
\sbox{\uuunit}
    {\setlength{\unitlength}{0.825em}
     \begin{picture}(0.6,0.7)
        \thinlines
        \put(0,0){\line(1,0){0.5}}
        \put(0.15,0){\line(0,1){0.7}}
        \put(0.35,0){\line(0,1){0.8}}
       \multiput(0.3,0.8)(-0.04,-0.02){10}{\rule{0.5pt}{0.5pt}}
     \end {picture}}

\makeatletter \@addtoreset{equation}{section} \makeatother

\def\bfone{\relax{\rm 1\kern-.35em 1}}

\def\bfone{\relax{\rm 1\kern-.35em 1}}
 
\def\downnormalfill{$\,\,\vrule depth4pt width0.4pt
\leaders\vrule depth 0pt height0.4pt\hfill\vrule depth4pt width0.4pt\,\,$}
\def\WT#1{\mathop{\vbox{\ialign{##\crcr\noalign{\kern3pt}
      \downnormalfill\crcr\noalign{\kern1.8pt\nointerlineskip}
      $\hfil\displaystyle{#1}\hfil$\crcr}}}\limits}

\begin{document}
\begin{titlepage}
\begin{flushright}
DFTT-2004\\
\end{flushright}
\vskip 1.5cm
\begin{center}
{\LARGE { \bf Tree Level Gravity - Scalar Matter Interactions}}

{\LARGE{\bf in Analogy with Fermi Theory of Weak Interactions}}

{\LARGE{\bf using Only a Massive Vector Field}}
%{\LARGE {\bf  Tree Level Interaction between Gravity and Scalar Matter in Analogy with  Fermi Theory of Weak Interactions }}
\vfill {\large
 Leonardo Modesto} \\
\vfill {
$^1$ Dipartimento di Fisica Teorica, Universit\'a di Torino, \\
$\&$ INFN -
Sezione di Torino\\
via P. Giuria 1, I-10125 Torino, Italy  }
\end{center}
\vfill
\begin{abstract}
{In this paper we
work in perturbative Quantum Gravity coupled to Scalar Matter at tree level and we introduce a new effective model in analogy with the Fermi theory of weak interaction and in relation with a previous work where we have studied only the gravity and its self-interaction. This is an extension of the I.T.B model (Intermediate-Tensor-Boson) for gravity also to gravitational interacting
scalar matter. We show that in a particular gauge the infinite series of interactions containing ``$n$'' gravitons and two scalars could be rewritten in terms of only two Lagrangians containing a massive field, the graviton and, obviously, the scalar field. Using the $S$-matrix we obtain that the low energy limit of the amplitude reproduce the local Lagrangian for the scalar coupled to gravity.}
\end{abstract}
%\vspace{2mm} \vfill \hrule width 3.cm {\footnotesize $^ \dagger $
%This work is supported in part by the European Union RTN contracts
%HPRN-CT-2000-00122 and HPRN-CT-2000-00131.}
\end{titlepage}
%%%%%%%%%%%%%%%%%%%%%%%%%%%%%%%%%%%%%%%%%%%%%%%%%%%%%%%%%%%%%%%%%%%%%%%%%%%
%\tableofcontents
%%%%%%%%%%%%%%%%%%%%%%%%%%%%%%%%%%%%%%%%%%%%%%%%%%%%%%%%%%%%%%%%%%%%%%%%%%%
\section{Introduction}
In this paper we consider the expansion of the action for a real scalar field coupled to gravity and we study the interaction between scalar particles and gravitons. In the Einstein gauge the interactions contain always two scalars and $1$, $2$, $\dots$ $n$ - gravitons. After the expantion we obtain an infinite number of local Lagrangian terms that contain at $n$th-order $n$-gravitons. In analogy with Fermi theory of weak interactions and its extension to the non-local I.V.B model (Intermediate Vector Boson) we introduce only two new Lagrangian terms and we reconstruct all the infinite interactions at tree level in the $S$-matrix at $n$-order. This model is an extension of the I.T.B. model \cite{ITB} to include the scalar matter. 
In our new model the interaction between $n$-gravitons and two scalar fields is non-local and it is 
mediated by a massive spin 1 particle. In the limit in which the mass of the particle goes to infinity we obtain the local interactions of the scalar theory in curved background. 
We call the model I.T.B (Intermediate Tensor Boson) as in the previous work, also if in this case the interaction is mediated by a vector particle.

\section{The I.T.B Model for the Scalar Field}
In this section we introduce the I.T.B model for a scalar field coupled to gravity. To do it we introduce two Lagrangian terms, and using only those two terms we reproduce the infinite interaction that contain "$n$"-gravitons and two scalar fields.  \\
The action for scalar field in curved space-time is 
\begin{equation}
S=\frac{1}{2}\int d^{4}x \sqrt{-g}[g^{\mu\nu} \partial_{\mu}\varphi\partial_{\nu}\varphi-m^{2}\varphi^{2}]
\label{eq451}
\end{equation}   
The gauge chosen is  (see the appendix) :
\begin{equation}
\partial_{\mu}\sqrt{-g}=0 \,\, \rightarrow \sqrt{-g}=const=1 \label{eq452}
\end{equation} 
Using the developed of the inverse metric $g^{\mu\nu}$ in powers of the graviton $h_{\mu \nu}$ ($g_{\mu \nu} = \eta_{\mu \nu} + h_{\mu \nu}$) we have the following form :
\begin{eqnarray} 
S&=&\frac{1}{2}\int d^{4}x[\sum_{n=0}^{\infty}(G_{N})^{\frac{n}{2}}(-1)^{n}(h^{n})^{\mu\nu}\partial_{\mu}\varphi\partial_{\nu}\varphi-m^{2}\varphi^{2}]= \nonumber \\
&=&\frac{1}{2}\int d^{4}x[\eta^{\mu\nu}\partial_{\mu}\varphi\partial_{\nu}\varphi-(G_{N})^{\frac{1}{2}}h^{\mu\nu}\partial_{\mu}\varphi\partial_{\nu}\varphi+(G_{N})h^{\mu}\,_{\sigma}h^{\sigma\nu}\partial_{\mu}\varphi\partial_{\nu}\varphi+ \nonumber \\
&&\hspace{1.6cm}-(G_{N})^{\frac{3}{2}}h^{\mu}\,_{\sigma}h^{\sigma}\,_{\rho}h^{\rho\nu}\partial_{\mu}\varphi\partial_{\nu}\varphi+\dots \dots -m^{2}\varphi^{2}]
\label{eq453}
\end{eqnarray}
We introduce the "$current$" 
\begin{equation}
J^{\mu}(x)=h^{\mu\sigma} \partial_{\sigma} \varphi(x)
\label{eq454}
\end{equation}
This current couples to a spin=1 massive particle as follow 
\begin{equation}
L_{(1)}=g_{1}J^{\mu}\Phi_{\mu}=g_{1}h^{\mu\sigma} \partial_{\sigma} \varphi \Phi_{\mu}
\label{eq455}
\end{equation}
This Lagrangian is sufficient to obtain a non local interaction for two gravitons and two scalars.
Our purpose is to obtain all interactions between "$n$"-gravitons and two scalar fields, therefore we introduce a second Lagrangian term that contains a graviton and two spin $1$ fields. 
 \begin{equation}
L_{(2)}=g_{2}M_{\Phi}h_{\mu\nu}\Phi^{\mu}\Phi^{\nu}
\label{eq456}
\end{equation}
Now we to calculate the tree level interaction between two gravitons an two scalars in the I.T.B model using the Lagrangian $L_{(1)}+L_{(2)}$, at the second order in the $S$-matrix
\begin{eqnarray}
S^{(2)}&=&\frac{(i)^{2}}{2} \int d^{4}x_{1}d^{4}x_{2}T[:L_{(1)}(x_{1})::L_{(2)}(x_{2}):]= \nonumber \\
&=&\frac{(i)^{2}}{2} \int d^{4}x_{1}d^{4}x_{2}(g_{1})^{2}[:(h_{\mu\sigma}\partial^{\mu}\varphi\WT{\Phi^{\sigma})(x_{1})::(h_{\alpha\beta}\partial^{\alpha}\varphi\Phi}\vphantom{.}^{\beta})(x_{2}):]= \nonumber \\
&=&\frac{(i)^{2}}{2} \int d^{4}x_{1}d^{4}x_{2}(g_{1})^{2}(h_{\mu\sigma}\partial^{\mu}\varphi)(x_{1}) \int \frac{d^{4}k}{(2 \pi)^{4}}(-i)\frac{\eta^{\sigma\beta}-\frac{k^{\sigma}k^{\beta}}{M_{\Phi}^{2}}}{k^{2}-M_{\Phi}^{2}} e^{ik(x_{2}-x_{1})} \times \nonumber \\
&&\hspace{6.4cm} \times (h_{\alpha\beta}\partial^{\alpha}\varphi\Phi^{\beta})(x_{2})\,+ \dots   \rightarrow \nonumber \\
&\rightarrow& (-i)\frac{(g_{1})^{2}}{2M_{\Phi}^{2}}\int d^{4}x\,(h_{\mu\sigma}\partial^{\mu}\varphi h^{\sigma}\,_{\alpha}\partial^{\alpha}\varphi)(x)+\dots
\end{eqnarray}
This is exactly the local amplitude that we obtain from the Lagrangian (\ref{eq453}), provided we make the following identification 
\begin{equation}
\frac{(g_{1})^{2}}{2M_{\Phi}^{2}}=\frac{G_{N}}{2}
\end{equation}
The S-matrix to $3^{rd}$- order is 
\begin{eqnarray}
S^{(3)}&=&\frac{(i)^{3}}{(3!)} \int d^{4}x_{1} d^{4}x_{2} d^{4}x_{3}  T[:(L_{(1)}+L_{(2)})(x_{1})::(L_{(1)}+L_{(2)})(x_{2}):\times \nonumber \\
&&\hspace{4.2cm}\times:(L_{(1)}+L_{(2)})(x_{3}): ]= \nonumber \\
&=&\frac{(i)^{3}}{(3!)} \int d^{4}x_{1} d^{4}x_{2} d^{4}x_{3}(g_{1}h_{\mu\sigma} \partial^{\mu}\varphi \WT{\Phi^{\sigma})_{1}(g_{2}M_{\Phi}h_{\alpha\beta}\Phi}\vphantom{.}^{\alpha}\WT{\Phi^{\beta})_{2}(g_{1}h_{\nu\rho} \partial^{\nu}\varphi \Phi}\vphantom{.}^{\rho})_{3} \nonumber \\
&\rightarrow& \frac{(i)}{(3!)}\times2\times3 \frac{g_{1}^{2}g_{2}}{M_{\Phi}^{4}}M_{\Phi}\int d^{4}x (h_{\mu\sigma}h^{\sigma}_{\rho}h^{\rho}_{\nu} \partial^{\mu}\varphi\partial^{\nu}\varphi)(x)= \nonumber \\
&=&(i)\frac{g_{1}^{2}g_{2}}{M_{\Phi}^{3}} \int d^{4}x (h_{\mu\sigma}h^{\sigma}_{\rho}h^{\rho}_{\nu} \partial^{\mu}\varphi\partial^{\nu}\varphi)(x)
\label{eq457}
\end{eqnarray}
In the $3^{rd}$ line factor $2$ comes from all possible contractions of the fields $\Phi$, factor $3$ from the fact that the Lagrangian is a sum of two terms and doing the product we obtain a sum of $3$ identical factors.\\
We continue the study of gravity coupled to scalar field going to study the $n^{th}$-order of the $S$-matrix. At this order the $S$-matrix reproduces the interactions between $n$-gravitons and two scalar fields
\begin{eqnarray}
S^{(n)}&=&\frac{(i)^{n}}{(n!)} \times [(n-2)!\,2^{n-2}] \times \frac{n(n-1)}{2} \int d^{4}x_{1} d^{4}x_{2} \dots d^{4}x_{n}\times \nonumber \\
&&\times T[:(L_{(1)}+L_{(2)})(x_{1})::(L_{(1)}+L_{(2)})(x_{2}): \dots :(L_{(1)}+L_{(2)})(x_{n}):]= \nonumber \\
&=&(i)^{n}\,2^{n-3}\int d^{4}x_{1} d^{4}x_{2} \dots d^{4}x_{n}\times \nonumber \\
&&\times T[:(L_{(1)}+L_{(2)})(x_{1})::(L_{(1)}+L_{(2)})(x_{2}): \dots :(L_{(1)}+L_{(2)})(x_{n}):]=\nonumber \\
&=&(i)^{n}\,2^{n-3}\int d^{4}x_{1} d^{4}x_{2} \dots d^{4}x_{n}\times \nonumber \\
&& \times (g_{1}h_{\mu\sigma} \partial^{\mu}\varphi \WT{\Phi^{\sigma})_{1}(g_{2}M_{\Phi}h_{\alpha\beta}\Phi}\vphantom{.}^{\alpha} \Phi^{\beta})_{2}\dots (g_{2}M_{\Phi}h_{\gamma\delta}\Phi^{\gamma}\WT{ \Phi^{\delta})_{n-1}(g_{1}h_{\epsilon\tau} \partial^{\epsilon}\varphi \Phi}\vphantom{.}^{\tau})_{n} \nonumber \\
\label{eq458}
\end{eqnarray}
In the previous equation the multipling term $[(n-2)!\,2^{n-2}]$ derives from all possible contractions of the massive fields $\Phi_{\mu}$, while the term $\frac{n(n-1)}{2}$ derives from the fact that the Lagrangian is a sum of two terms and so calculating the product between Lagrangians  
defined in different points we obtain more copies of the same amplitude.\\
%The amplitude graph is in the figure.\\
Now we compare the result obtained from amplitude \ref{eq458} with the local Lagrangian (\ref{eq453}). We obtain the relation
\begin{equation}
\frac{2^{n-3}}{M_{\Phi}^{n}}g_{1}^{2}g_{2}^{n-2}=\frac{G_{N}^{\frac{n}{2}}}{2}
\end{equation}
that is valid to all orders in the $S$-matrix. \\
If we redefine 
$g_{2}\,\rightarrow\,\frac{g_{2}}{2}$, we obtain 
\begin{equation} 
\frac{g_{1}^{2}g_{2}^{n-2}}{M_{\Phi}^{n}}=G_{N}^{\frac{n}{2}}
\label{eq460}
\end{equation}
For $n=2$ ed $n=3$ we have 
\begin{eqnarray} 
&\frac{g_{1}^{2}}{M_{\Phi}^{2}}=G_{N}=\frac{1}{M_{P}^{2}}\\
&\frac{g_{1}^{2}g_{2}}{M_{\Phi}^{3}}=G_{N}^{\frac{3}{2}}=\frac{1}{M_{P}^{3}}
\label{eq461}
\end{eqnarray}
and from that :
\begin{eqnarray} 
&\frac{g_{1}}{M_{\Phi}}=\frac{1}{M_{P}}\\
&\hspace{0.4cm}g_{1}=g_{2}
\label{eq462}
\end{eqnarray}
The question now is if the theory fixes a limit for the mass of the spin $1$ field.\\
The Lagrangian $L_{(1)}$ has adimensional coupling constant that can be very small. 
The Lagrangian $L_{(2)}$ instead has a coupling constant with mass dimension  
$g_{2}M_{\Phi}$. Assuming 
\begin{eqnarray} 
&g_{1}=g_{2} \equiv g \ll 1\\
&\frac{g_{1}}{M_{\Phi}}=\frac{1}{M_{P}}
\label{eq463}
\end{eqnarray}
Than 
\begin{equation}
M_{\Phi}\ll M_{P}
\end{equation}
and if $g_{2}\sim 10^{-15}$, we obtain the following mass for $\Phi$ :
\begin{equation} 
M_{\Phi}\sim 10^{4}GeV
\label{eq465}
\end{equation}
It is open therefore the possibility to observe quantum gravitational effects at lower energy than Planck scale. In the I.T.B model doesn't  exist the Planck scale because the $G_N$ constant has become the mass of a spin 1 particle. It is analog to the I.V.B model of weak interactions 
where the Fermi constant of weak interactions becomes proportional to the mass of the mediators $W^{+}$, $W^{-}$.
%Nel modello appena studiato il campo scalare $\varphi$ pu\`o essere il bosone di Higgs per cui, se $M_{\Phi} \sim 10^{4}-10^{5}\, GeV$, possiamo osservare negli acceleratori un suo accoppiamento alla gravit\`a ad energie uguali o poco superiori alla massa dell'Higgs. Torno a sottolineare che nelle nostre ipotesi $M_{\Phi} \ll M_{P}$ e quindi possiamo ad esempio prendere $M_{\Phi} \sim 10^{4}\,GeV$.
\section{Outlook and conclusions}
In this paper we introduced the I.T.B Model (Intermediate Tensor Boson) \cite{ITB} for gravity coupled to scalar matter. In this case in the particular suitable gauge $\sqrt{- g} = const$ we  expressed all the infinity tree level amplitudes for gravity in the presence of scalar matter using only two Lagrangian terms. The first Lagrangian term contains a graviton field, a scalar and a vector massive field. The second Lagrangian term contains two spin one massive fields and a graviton. We can write the complete interaction Lagrangian for the I.T.B model as following 
\begin{equation}
L_I = g_{1}h^{\mu\sigma} \partial_{\sigma} \varphi \Phi_{\mu} + g_{2}M_{\Phi}h_{\mu\nu}\Phi^{\mu}\Phi^{\nu} - \frac{g_1}{2 M_{\Phi}} h^{\mu \nu}\partial_{\mu} \phi \, \partial_{\nu} \phi 
\end{equation}
We repeat that only with this interaction Lagrangian we can reproduce all the infinite interaction terms for scalar matter interacting with gravity present in the action (\ref{eq453}).\\
In addiction we obtain, such in the case of I.T.B model for pure gravity, that the square root of the Newton constant $G_N$ is proportional to the inverse of mediator mass $G_N \sim 1/M_{\Phi}^2$.    
\section*{Acknowledgements}
We are grateful to Enore Guadagnini, Eugenio Bianchi, Gabriele Marchi and Giuseppe Tarabella for many important and clarifying discussions.
\newpage
\appendix
\section{The Gauge $\sqrt{(-g)}= const$}
In the I.T.B model for the scalar field we have introduced the gauge $\sqrt{(-g)}=const $. Now we to study this gauge.  This gauge is 
\begin{equation}   
\Gamma^{\lambda}_{\mu\lambda}=\frac{1}{2}g^{\lambda\sigma}(\partial_{\mu}g_{\lambda\sigma})=\frac{1}{\sqrt{(-g)}}\partial_{\mu}\sqrt{(-g)}=0
\label{eq466}
\end{equation}
We expand $g_{\mu\nu}$ near the flat metric and we check that it is always locally possible using  a diffeomorfism to go from a generic metric to a new metric which satisfies it. \\
To first order in $h_{\mu\nu}$ the gauge becomes 
\begin{eqnarray}
&&\frac{1}{2}\eta^{\lambda\sigma}(\partial_{\mu}h_{\lambda\sigma})=0\nonumber \\
&&\partial_{\mu}h^{\sigma}\,_{\sigma}=0
\label{eq467}
\end{eqnarray}
Under an infinitesimal coordinate transformation the fluctuation transforms as follows 
\begin{eqnarray}  
&&x^{\prime}_{\mu}=x_{\mu}+\varepsilon_{\mu} \nonumber \\
&&h^{\prime}_{\mu\nu}=h_{\mu\nu}+\partial_{\mu}\varepsilon_{\nu}+\partial_{\nu}\varepsilon_{\mu}
\label{eq468}
\end{eqnarray}
Using  \ref{eq468} in \ref{eq467}, we find 
\begin{eqnarray}  
&&\frac{1}{2}\eta^{\lambda\sigma}(\partial_{\mu}h^{\prime}_{\lambda\sigma})=\frac{1}{2}\eta^{\lambda\sigma}(\partial_{\mu}h_{\lambda\sigma})+\frac{1}{2}\eta^{\lambda\sigma}\partial_{\mu}(\partial_{\lambda}\varepsilon_{\sigma}+\partial_{\sigma}\varepsilon_{\lambda})\nonumber \\
&&0=\partial_{\mu}h^{\sigma}\,_{\sigma}+\partial_{\mu}\partial^{\sigma}\varepsilon_{\sigma} \nonumber \\
&&\partial^{\sigma} \varepsilon_{\sigma}+h^{\sigma}\,_{\sigma}=\, cost.
\label{eq469}
\end{eqnarray}
Now we define 
\begin{equation}
\varepsilon_{\sigma}\equiv\partial_{\sigma}\varphi(x)  
\label{eq470}
\end{equation}
Introducing \ref{eq470} in \ref{eq469} we get
\begin{equation}
\Box  \varphi = cost.-h^{\sigma}\,_{\sigma}
\label{eq471}
\end{equation}
The d'Alambert operator is invertible on $L_2$ functions, so the solution of \ref{eq471} is  
\begin{equation} 
\varphi(x)=-\frac{1}{\Box}(cost.-h^{\sigma}\,_{\sigma})
\label{eq472}
\end{equation}
Therefore the condition \ref{eq466} is attainable with a diffeomorfism.\\
To preserve the causality we introduce the Feynman prescription to define the inverse d'Alambertian operator :
\begin{equation}
\Box\, \rightarrow \, \Box+i\epsilon
\label{eq473}
\end{equation} 
Introducing the prescription \ref{eq473}, the solution of  \ref{eq471} is 
\begin{equation}
\varphi(x)=-\int \frac{d^{4}p}{(2\pi)^{4}} \frac{e^{-ipx}}{p^{2}+i\epsilon}[(2\pi)^{4}\delta(p)\times cost.-(\mathcal Fh)^{\sigma}\,_{\sigma}(p)]
\label{eq474}
\end{equation}
Where $(\mathcal{F}h)^{\sigma}\,_{\sigma}$ denote the Furier Transform of $h^{\sigma}\,_{\sigma}$.\\
The Lagrangian for the free graviton in this gauge is 
\begin{equation}
L=\partial_{\mu}h_{\nu\rho}\partial^{\mu}h^{\nu\rho}-2\partial_{\mu}h_{\nu\rho}\partial^{\nu}h^{\mu\rho}
\label{eq475}
\end{equation}
While the equation of motion becomes 
\begin{equation}
\Box h_{\mu\nu}-\partial_{\mu}\partial^{\rho}h_{\rho\nu}-\partial_{\nu}\partial^{\rho}h_{\rho\mu}=0
\label{eq476}
\end{equation}
The residual gauge invariance (of parameter $\varepsilon(x)$) of \ref{eq476} is 
\begin{equation} 
\Box h^{\prime} _{\mu\nu}-\partial_{\mu}\partial^{\rho}h^{\prime} _{\rho\nu}-\partial_{\nu}\partial^{\rho}h^{\prime} _{\rho\mu}=\Box h_{\mu\nu}-\partial_{\mu}\partial^{\rho}h_{\rho\nu}-\partial_{\nu}\partial^{\rho}h_{\rho\mu}-2\partial_{\mu}\partial_{\nu}(\partial^{\rho}\varepsilon_{\rho})
\label{eq480}
\end{equation}
To obtain invariance in \ref{eq480} we must take the parameter to satisfy   
\begin{equation} 
\partial_{\mu}(\partial^{\sigma}\varepsilon_{\sigma})=cost.
\end{equation}
but to preserve the gauge condition $\ref{eq469}$ we must take $const = 0$.

%and this is exactly the equation  \ref{eq469}, in fact at infinitesimal level
%the gauge condition became $\partial_{\mu}h^{\nu}\,_{\nu}=0$.


\begin{thebibliography}{99}
%
\bibitem{ITB} Leonardo Modesto \emph{Perturbative Quantum Gravity in Analogy with Fermi Theory of Weak   Interactions using Bosonic Tensor Fields} [hep-th/0312318]
 %
\bibitem{SF0} A. Sagnotti and M. Tsulaia \emph{On higher spins and the tensionless limit of String Theory}, [hep-th/0311257]
%
\bibitem{SF1} Dario Francia and Augusto Sagnotti \emph{Free geometric equations for higher spins}, Phys. Lett. B {\bf 543} (2002) 303 [hep-th/0207002]; 
Dario Francia and Augusto Sagnotti 
\emph{On the geometry of higher-spin gauge fields}, Class. Quant. Grav. {\bf 20} (2003) S473 [hep-th/0212185]
%
\bibitem{HSF} 
W.~Siegel and B.~Zwiebach,
%``Gauge String Fields,''
Nucl.\ Phys.\ B263 (1986) 105;
%%CITATION = NUPHA,B263,105;%%
T.~Banks and M.~E.~Peskin,
%``Gauge Invariance Of String Fields,''
Nucl.\ Phys.\ B264 (1986) 513
%
\bibitem{M01} C.~Fronsdal,
%``Massless Fields With Integer Spin,''
Phys.\ Rev.\ D18 (1978) 3624;
%%CITATION = PHRVA,D18,3624;%%
T.~Curtright,
%``Massless Field Supermultiplets With Arbitrary Spin,''
Phys.\ Lett.\ B85 (1979) 219;
%
%\bibitem{M02} 
J.~Fang and C.~Fronsdal,
%``Massless Fields With Half Integral Spin,''
Phys.\ Rev.\ D18 (1978) 3630
\bibitem{VasilevEq}M.~A.~Vasiliev,
%``Consistent Equation For Interacting Gauge Fields Of All Spins In (3+1)-Dimensions,''
Phys.\ Lett.\ B243 (1990) 378,
%%CITATION = PHLTA,B243,378;%%
%``Properties Of Equations Of Motion Of Interacting Gauge Fields Of All Spins In (3+1)-Dimensions,''
Class.\ Quant.\ Grav.\  8 (1991) 1387,
%%CITATION = CQGRD,8,1387;%%
%``Algebraic Aspects Of The Higher Spin Problem,''
Phys.\ Lett.\ B257 (1991) 111,
%%CITATION = PHLTA,B257,111;%%
%``More on equations of motion for interacting massless fields of all spins in (3+1)-dimensions,''
Phys.\ Lett.\ B285 (1992) 225, 
Int.\ J.\ Mod.\ Phys.\ D5 (1996) 763
[arXiv:hep-th/9611024],
%%CITATION = HEP-TH 9611024;%%
%``Higher-Spin-Matter Gauge Interactions In 2+1 Dimensions,''
Nucl.\ Phys.\ Proc.\ Suppl.\  56B (1997) 241,
%%CITATION = NUPHZ,56B,241;%%
%``Higher spin gauge theories: Star-product and AdS space,''
%arXiv:hep-th/9910096,
%%CITATION = HEP-TH 9910096;%%
%``Progress in higher spin gauge theories,''
arXiv:hep-th/0104246.
\bibitem{200}
B.~de Wit and D.~Z.~Freedman,
%``Systematics Of Higher Spin Gauge Fields,''
Phys.\ Rev.\ D21 (1980) 358;
%%CITATION = PHRVA,D21,358;%%
T.~Damour and S.~Deser,
%``'Geometry' Of Spin 3 Gauge Theories,''
Annales Poincare Phys.\ Theor.\  {\bf 47} (1987) 277.
%
\bibitem{300} Emidio Gabrielli \emph{Extended Pure Yang - Mills Gauge Theories with Scalar and Tensor Gauge Field}  Phys. Letters {\bf{B}} 258,151-155 (1991); \emph{Extended Gauge Theories in Euclidean Space with Higher Spin Fields} Annals Physics 287, 229-259 (2001) [hep-th/9909117]
\bibitem{400} D. G. C. McKeon, T.N. Sherry \emph{Gauge Model with Extended Field Transformations in Euclidean Space}, Int. J. Mod. Physics {\bf A} 15, 227-250 (2000) [hep-th/9811102]  
\bibitem{TRZ1}
K. Akama, Y. Chikashige, T. Matsuki and H. Terazawa  \emph{Gravity and Electromagnetism as Collective Phenomena of Fermion - Antifermion Pairs}, Progress of Theoretical Physics, Vol. 60, No. 3, September 1978; \\
H. Terazawa  \emph{Supergrand Unification of Gravity with all the Other Fundamental Forces}, Institute for Nuclear Study Universiy of Tokio, INS - Rep. - 338 July, 1979; \\
H. Terazawa and K, Akama  \emph{Dynamical Subquark Model of Pregauge and Pregeometric Interactions}, Phys. Letters Vol. 96 {\bf B} No. 3, 4 , 276, 3 November 1980; \\ 
H. Terazawa  \emph{Pregeometry}, Institute for Nuclear Study Universiy of Tokio, INS - Rep. - 429 October, 1981; \\
D. Amati and G. Veneziano \emph{A Unified Gauge and Gravity Theory with only Matter Fields}, 
Nuclear Physics {\bf B} 204 (1982) 451; \\ 
 H. Terazawa  \emph{A Fundamental Theory of Composite Particle and Fields}, Institute for Nuclear Study Universiy of Tokio, INS - Rep. - 463 Feb. , 1983; \\
H. Terazawa  \emph{Sepergrand Unified Composite Model in Pregeometry} Institute for Nuclear Study Universiy of Tokio, INS - Rep. - 511 Oct , 1984
%
\bibitem{TRZ2}
K. Akama, Y. Chikashige and H. Terazawa  \emph{What are the Gauge Bosons made of?} Institute for Nuclear Study Universiy of Tokio, INS - Rep. - 261 June, 1976
%
\bibitem{WETT}
A. Hebecker and C. Wetterich \emph{Spinor Gravity}, Phys.Lett. {\bf{B}} 574 (2003) 269-275, [hep-th/0307109]; \\
C. Wetterich \emph{Gravity from Spinors}, [hep-th/0307145]
\end{thebibliography}
\end{document}